# Core Electron Heating By Triggered Ion Acoustic Waves In The Solar Wind

By F.S. Mozer[1,2], S.D. Bale[1,2], C.A. Cattell[3], J. Halekas[4], I.Y. Vasko[1], J.L. Verniero[5] and P.J. Kellogg[3]
  1. Space Sciences Laboratory, University of California, Berkeley, USA
  2. Physics Department, University of California, Berkeley, USA
  3. University of Minnesota, Minneapolis, Mn, USA
  4. University of Iowa, Iowa City, Iowa, USA
  5. Goddard Space Flight Center, Greenbelt, Md, USA

Perihelion passes on Parker Solar Probe orbits six through nine have been studied to show that solar wind core electrons emerged from 15 solar radii with a temperature of 55±5 eV, independent of the solar wind speed which varied from 300 to 800 km/sec. After leaving 15 solar radii and in the absence of triggered ion acoustic waves at greater distances, the core electron temperature varied with radial distance, R, in solar radii, as $1900R^{-4/3}$ electron volts because of cooling produced by the adiabatic expansion. The coefficient, 1900, reproduces the minimum core electron perpendicular temperature observed during the 25 days of observation. In the presence of triggered ion acoustic waves, the core electrons were isotropically heated as much as a factor of two above the minimum temperature, $1900R^{-4/3}$ eV. Triggered ion acoustic waves were the only waves observed in coincidence with the electron core heating. They are the dominant wave mode at frequencies greater than 100 Hz at solar distances between 15 and 30 solar radii.

**Introduction**

Ion acoustic waves have been observed by many satellites in the solar wind [Gurnett and Anderson, 1977; Gurnett and Frank, 1978; Kurth et al, 1979; Lin et al, 2001; Mozer et al, 2020A; Pisa et al, 2021]. They are wideband, short duration waves which differ greatly from the triggered ion acoustic waves seen at 15-30 solar radii on the Parker Solar Probe (PSP) [Mozer et al, 2021], in that these latter waves are narrowband, long duration (hours to days) waves that often appear as shock-like bursts of 100-1000 Hz waves whose bursts are phase-locked with low frequency ion-acoustic-like waves. The purpose of this paper is to study the effects of these triggered ion acoustic waves on the electron plasma. It has long been known that the core solar wind electrons must be heated as they move away from the Sun [Hartle and Sturrock, 1968] to overcome some of the temperature loss associated with their adiabatic expansion, but the heating mechanism has not previously been identified in experimental data. The role of ion acoustic waves in this heating has been discussed theoretically [Dum, 1978] and via calculations [Kellogg, 2020]. This paper will further examine their role as a mechanism for electron heating, thereby addressing the PSP mission science goal, to "trace the flow of energy that heats…the solar corona and solar wind" [Fox, 2016]. The Fields [Bale et al, 2016; Malaspina, 2016; Mozer et al, 2020B] and SWEAP [Halekas et al, 2020; Kasper et al, 2016; Whittlesey et al, 2020] instruments on the Parker Solar Probe obtained the data presented in this paper.

A gas expanding into the void cools as it expands. To derive an expression for this cooling, it is assumed that $pV^{\gamma}$ is constant, where $P = nkT_E$ is the plasma pressure, n is the density, V is the gas volume and $\gamma = 5/3$ for a monatomic gas. Since $n \propto R^{-2}$ and $V \propto R^2$, where R is the distance from the Sun, these equations combine to yield a core electron perpendicular temperature that varies with radius as $R^{-4/3}$ in the absence of in situ heating. Orbits six through nine were fit to a radial profile of the form $1900R^{-4/3}$ eV, where the coefficient, 1900, produced the minimum temperature observed on the 25 observation days. In the following discussion, the observed perpendicular temperature will be compared with this minimum temperature to determine locales where electron heating may have occurred. Other than in figure 5, all figures give only the core perpendicular temperature (and not the parallel temperature) because the two temperatures are essentially equal (see figure 5) so the parallel temperature is omitted for clarity in the figures.

**Data**

Figure 1 presents data from orbits 6 and 8 perihelia. Panels 1A and 1G give the electric field spectra, panels 1B and 1H give the core perpendicular electron temperature in black and the minimum temperature, $1900R^{-4/3}$ eV, in red, panels 1C and 1I give the ion perpendicular temperature, panels 1D and 1J give the solar wind speed, panels 1E and 1K give $T_E/T_I$, and panels 1F and 1L give the spacecraft distance from the Sun in units of solar radii. Halekas et al {2020} provided the electron temperature data shown throughout this paper.

As seen in panel 1A, there were two types of wave activity, the broadband, short duration, normal ion acoustic waves that appear as vertical lines and the narrow band, longer duration waves at the beginning of the interval, that lasted for times ~1 hour, and that are the triggered ion acoustic waves. Expanded examples of each of these waves are given in figure 2. Panel 2A illustrates the spectrum of the triggered ion acoustic waves that were present at early times in figure 1A. As shown, they are narrowband waves that existed in this case for times the order of an hour. They occurred in regions where $T_E/T_I$ was large (panel 2B) and the electron temperature of panel 2C exceeded the minimum temperature, $1900R^{-4/3}$ eV, by more than a factor of two. The one minute of data in panel 2D illustrates the normal ion acoustic waves that were broadband structures lasting for times the order of seconds.

The electron temperature of panel 1B exceeded the minimum temperature curve of $1900R^{-4/3}$ for two reasons. One reason was the 10-20% temperature increases due to the weak anti-correlation of the temperature with the solar wind speed during days 26 through 29 of figure 1B when the solar wind speed varied from 300 to 800 km/sec and the only observed waves were the normal ion acoustic waves. The second reason was the greater than a factor of two temperature increase in the first day of data in panel 1B in the presence of triggered ion acoustic waves, also illustrated in figures 2A and 2C.

The dependencies of the electron temperature described above also occurred during orbit eight in panels 1G-1L. Early in the interval and at occasional later times, the temperature of panel 1H was near the red curve in the absence of triggered ion acoustic waves and in the presence of small $T_E/T_I$, while, at other times, the temperature greatly exceeded the red minimum temperature curve during periods of triggered ion acoustic waves. It is reminded that the red curve describes the coldest electron temperature observed during the two passes in figure 1, as well as in figure 3 for perihelia 7 and 9. Thus, the summary of the data in figures 1 and 3 is that, in the absence of triggered ion acoustic waves, the electron temperature was greater than the minimum temperature by ~10-20% while it exceeded the minimum temperature by as much as a factor of two during periods of triggered ion acoustic waves.

Figures 4A-4H provide expanded views of two intervals between 21 and 30 solar radii in orbit 7 when the triggered ion acoustic waves (and some normal ion acoustic waves) were observed. Looking from perihelion outward in either direction, the red curves in Figure 4C and 4G show the minimum temperature decrease expected from adiabatic expansion. After perihelion, in panel 4G, the temperature further increased above the minimum temperature to signify that there was a significant core electron perpendicular temperature increase (the black curve in panel 4G) at the time of the triggered ion acoustic wave in panel 4E. Before perihelion, in panel 4C, there is a smaller temperature increase at the time of the triggered ion acoustic waves. Each of these curves are one-hour running averages made from about 500,000 raw data points.

Mixing of spatial and temporal variations is illustrated in figures 4A and 4C, where heated electrons were also observed at lower altitudes (later times) when there were no triggered ion acoustic waves. This situation can result from electrons heated by waves at a lower altitude than the satellite location such that the heated electrons arrived at the spacecraft before or in the absence of the slower moving waves. Thus, one cannot expect a detailed one-to-one correlation between the waves and the heating but there should be a general correlation, such as that shown during all of the heating events.

Figure 5 gives the electron and ion temperatures versus radial distance, obtained from combining the data of orbits six through nine into a single file as a function of radial distance and performing a one-hour running average of this data.  The left panel shows that core electron heating occurred between 20 and 25 solar radii because the core perpendicular and parallel electron temperatures rose above the minimum temperature red curve between 20 and 25 solar radii and they remained above that curve at larger radial distances. Each plot consists of some 10,000,000 raw data points.

Because the ion temperature at 15 solar radii was strongly dependent on the solar wind speed (see figures 1 and 3), it is not possible to determine the ion heating as a function of radial distance from the available data (right panel of figure 5) without correction of the ion temperature dependence on the solar wind speed.

Whistler waves are not seen in the 20-25 solar radius region [Cattell, et al, 2021].

Triggered ion acoustic waves are distinguished from normal ion acoustic waves by their narrowband frequency and longer duration, which may be understood [Kellogg, 2021]. Very often, but not always, the triggered ion acoustic waves have the following properties (see figure 6), which are not understood theoretically;
1. A few Hz electric field wave is present, accompanied with bursts of a few hundred to 1000 Hz waves whose bursts are phase locked with each low frequency period (figure 6A).
2. These periodic field structures can exist for large fractions of a day (see Figure 4 of Mozer et al [2021]).
3. Few Hz and few hundred Hz plasma density fluctuations (determined from the spacecraft potential) are present at the two frequencies (figures 6B and 6C).
4. No magnetic field signature is observed at either of these two wave frequencies (figure 6D).
5. The higher frequency electric field and density fluctuations are pure sine waves, indicating that the waves are very narrowband (figures 6E and 6F).

**Discussion**

The summary, from the statistics available on the altitude range of 15-30 solar radii during four orbits (eight passes, three of which are shown in detail in figures 2 and 4), is:
1. There were two passes with $(T_E/T_I) \leq 1$, few waves, and no electron heating.
2. There were six passes with $(T_E/T_I) > 1$, with triggered ion acoustic waves, and with correlated core electron heating.

The conclusions drawn from this limited data set are:
1. In the absence of triggered ion acoustic waves, the core electron perpendicular temperature was 10-20% greater than the minimum temperature of $1900R^{-4/3}$ (51-61 eV at 15 solar radii) in spite of solar wind speed variations from 300 to 800 km/sec.
2. In the presence of triggered ion acoustic waves, the electron temperature increased by as much as a factor of two above the minimum temperature, $1900R^{-4/3}$ eV.

These findings are consistent with the hypothesis that triggered ion acoustic waves contribute to core electron heating in the inner heliosphere.

Future PSP orbits will investigate the triggered ion acoustic waves to provide further statistics on their occurrence and their electron heating. They will also provide higher frequency measurements of plasma density fluctuations (through measurements of the spacecraft potential), which may be a significant contributor to electron heating and to achieving the PSP mission goal of assessing the dominant mechanisms of energy exchange and plasma heating near the Sun.

**Acknowledgements**

This work was supported by NASA contract NNN06AA01C. The authors acknowledge the extraordinary contributions of the Parker Solar Probe spacecraft engineering team at the Applied Physics Laboratory at Johns Hopkins University. The FIELDS experiment on the Parker Solar Probe was designed and developed under NASA contract NNN06AA01C. Our sincere thanks to P. Harvey, K. Goetz, and M. Pulupa for managing the spacecraft commanding, data processing, and data analysis, which has become a heavy load thanks to the complexity of the instruments and the orbit. We also acknowledge the SWEAP team for providing the plasma data. The work of I.V. was supported by NASA Heliophysics Guest Investigator grant 80NSSC21K0581

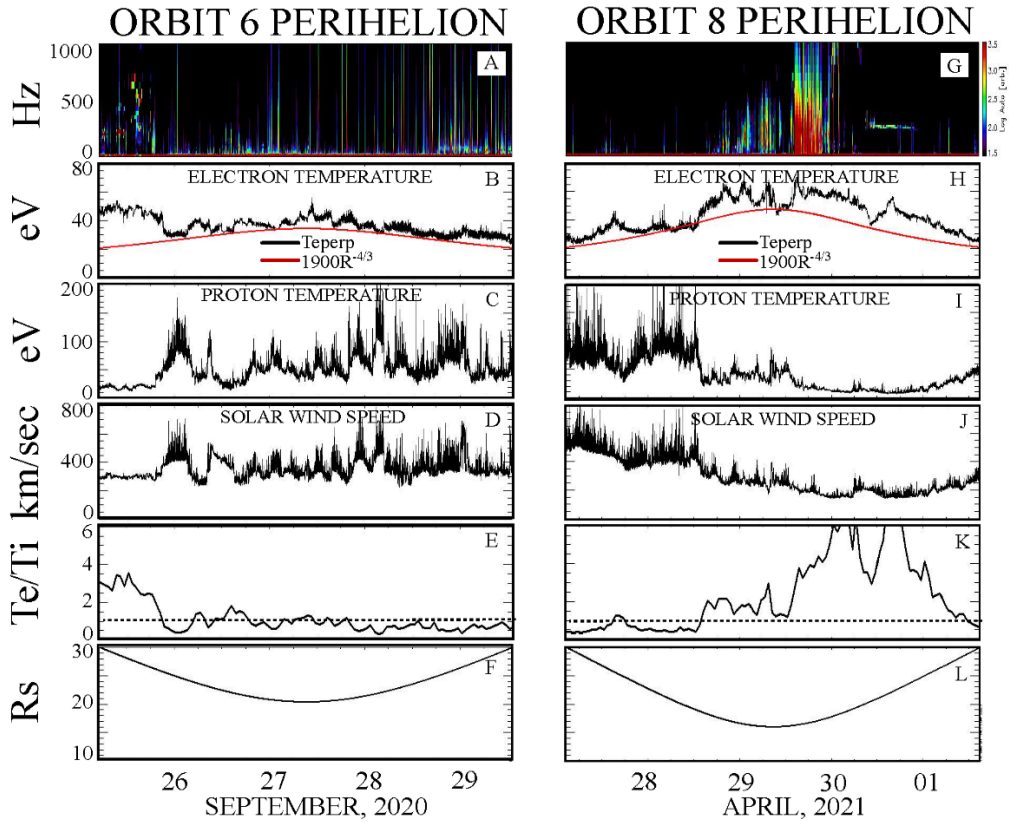

Figure 1. Plasma parameters during perihelion passes six and eight of the Parker Solar Probe. Note that after the first day of orbit six, the electron temperature followed the minimum temperature red curve even though the solar wind speed varied from 300 to 800 km/sec.

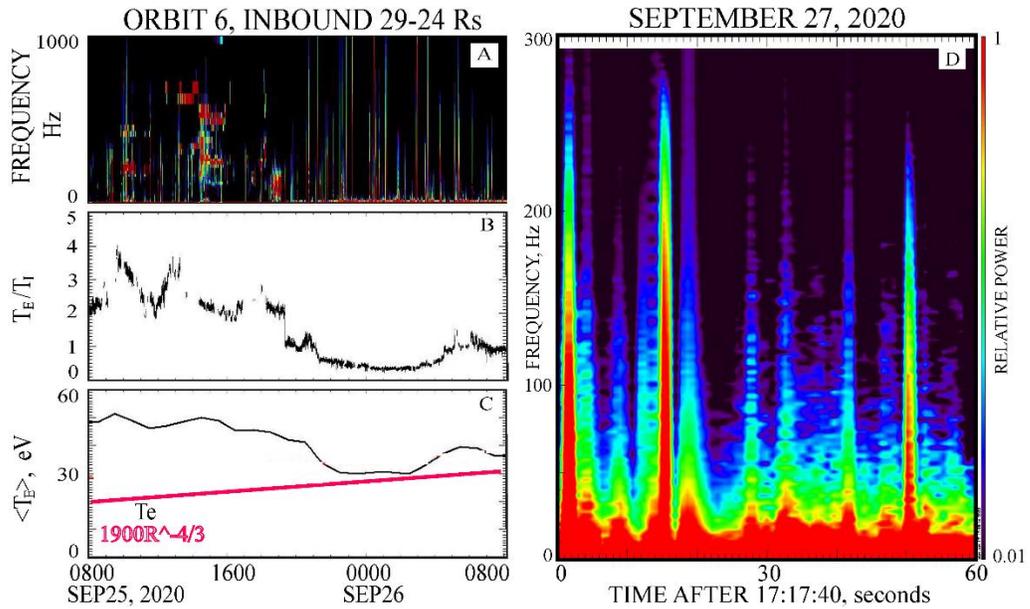

Figure 2. Triggered and normal ion acoustic waves observed during orbit 6. The triggered ion acoustic waves (panel 2A) were associated with large $T_E/T_I$ (panel 2B) and heated electrons (panel 2C). Note that the triggered ion acoustic waves were narrowband in frequency and long duration (hours long) while the normal ion acoustic waves during the one minute interval of panel 2D were broadband, short duration waves.

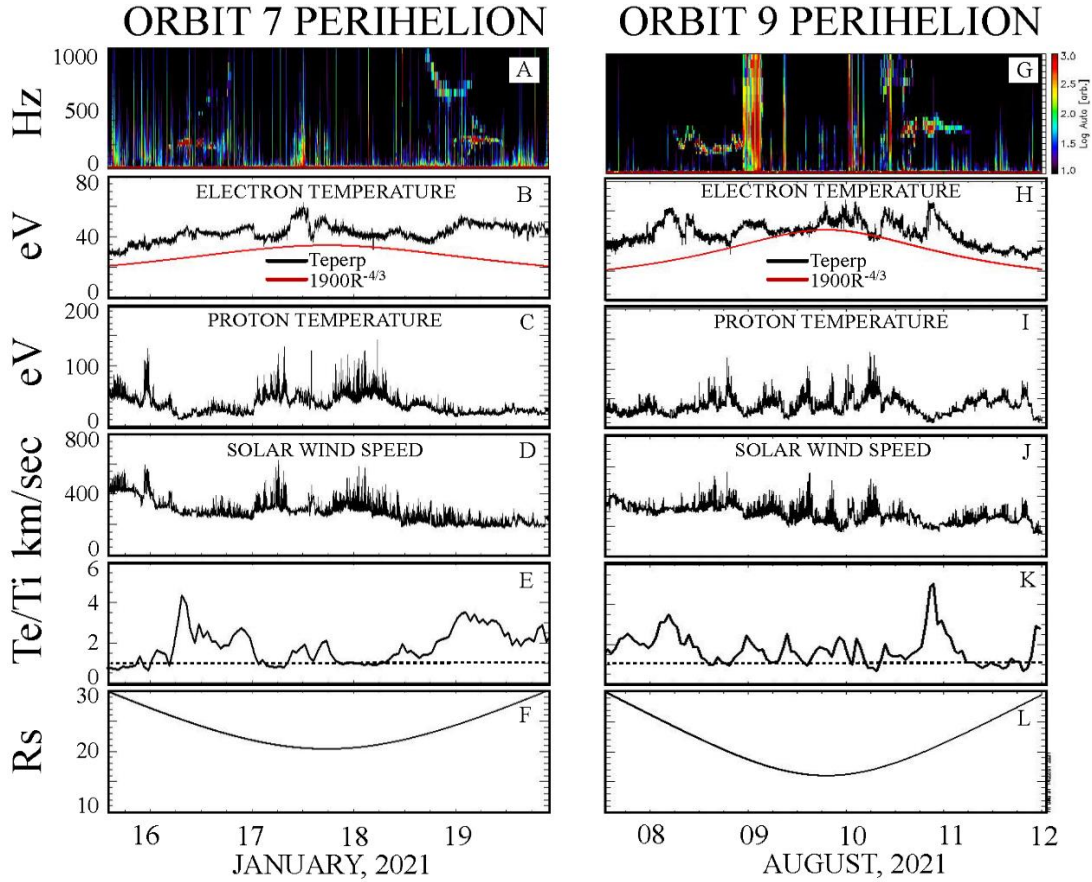

Figure 3. Plasma parameters during perihelion passes seven and nine of the Parker Solar Probe. Note that the electron temperature was approximately equal to but not less than the minimum temperature red curve in the absence of waves and that it exceeded the red curve by as much as a factor of two near triggered ion acoustic waves.

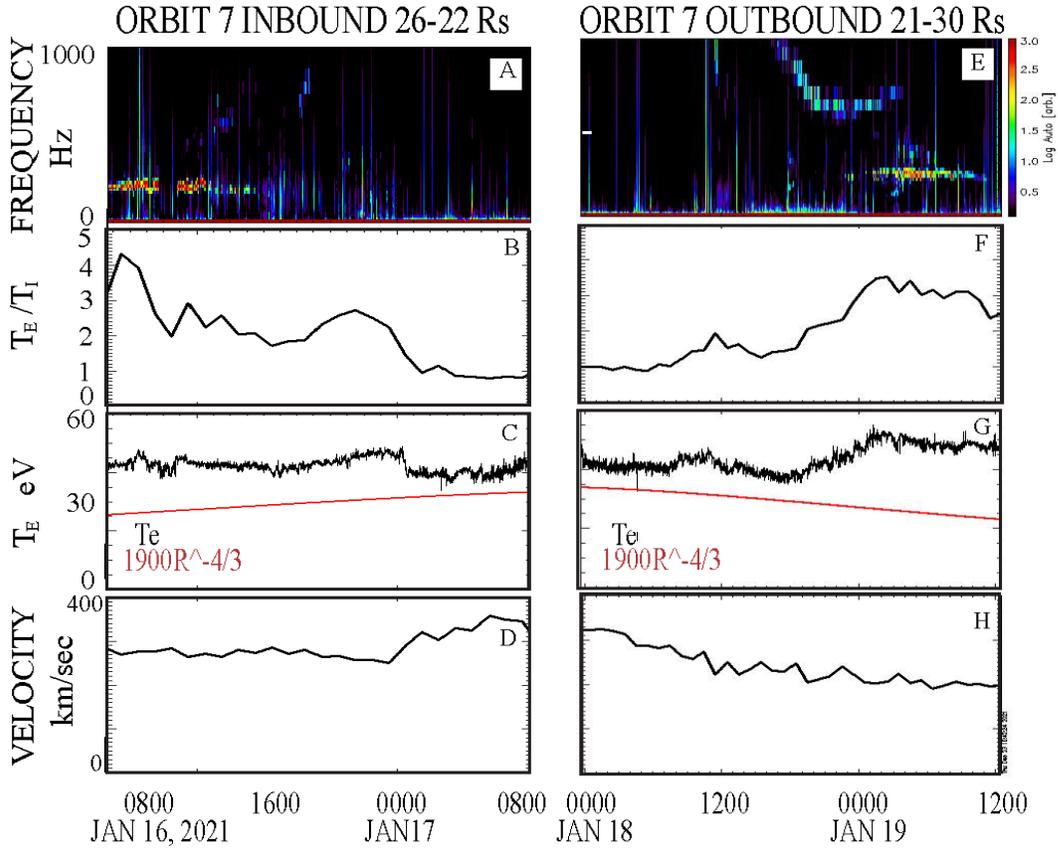

Figure 4. The inbound (panels 4A-4D) and outbound (panels 4E-4H) passes of orbit 7, illustrating the electric field, temperature ratio, electron temperature and the plasma velocity. The minimum temperature red curves in panels 4C and 4G give the electron temperature decreases with distance expected for adiabatic expansion of the electron gas without heating. Their differences from the measured temperatures (the black curves) show that electrons were heated in regions containing triggered ion acoustic waves (panels 4A and 4E) and large electron to ion temperature ratios (panels 4B and 4F).

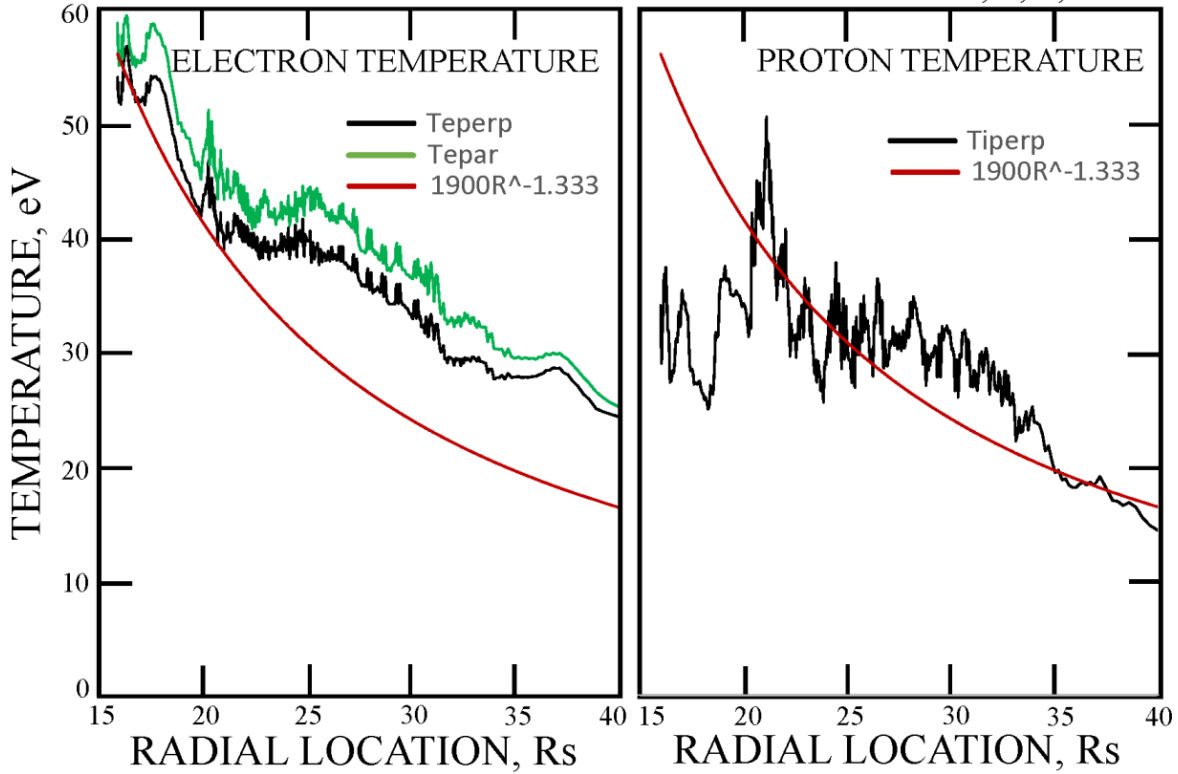

Figure 5. The core electron and ion temperatures versus solar radius, as determined from the eight passes of the Parker Solar Probe through the 15-40 solar radius region on orbits six through nine. Each curve gives the one-hour running average of some 10,000,000 raw data points. The red minimum temperature curves give the radial variation of the temperature for adiabatic expansion with no heating. The deviation between the electron core temperatures and the red curve in the left panel shows that electrons were heated in the 20-25 solar radius region. Because the ion temperature depended much more strongly on the solar wind speed than did the electron temperature, it is not possible to determine the ion heating in the right panel without correcting for the dependence of the ion temperature on the solar wind speed.

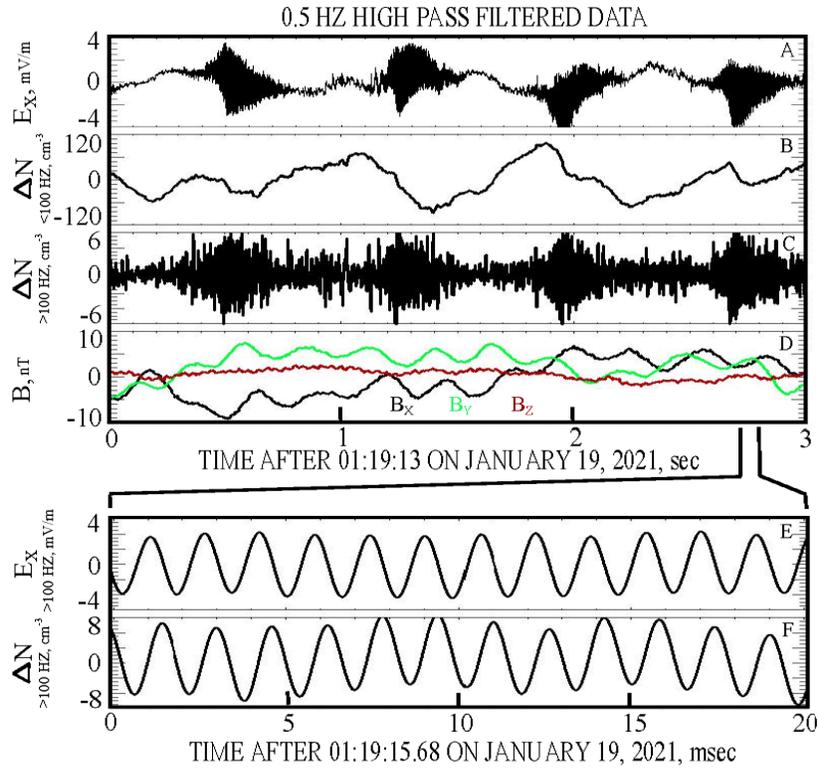

Figure 6. Features of triggered ion acoustic waves, including the electric field having higher frequency pulses in phase with a few Hz wave (Panel 6A), low and high frequency density fluctuations in sync with the electric fields (panels 6B and 6C), the absence of similar frequency magnetic field signatures (panel 6D), and with the higher frequency electric field and density fluctuations being narrowband signals (panels 6E and 6F).